\documentclass[
superscriptaddress,
amsmath,amssymb,
 ps,
showpacs,
prb,
]{revtex4-2}

\usepackage{newtxtext,newtxmath}

\usepackage{graphicx}
\usepackage{dcolumn}
\usepackage{bm}

\begin{document}

\title{Competitive coexistence of ferromagnetism and metal--insulator transition of VO$_2$ nanoparticles}

\author{Tsuyoshi Hatano}
\affiliation{College of Engineering, Nihon University, 1 Nakagawara, Tokusada, Tamura, Koriyama, Fukushima, 963-8642, Japan} 

\author{Akihiro Fukawa}
\affiliation{College of Science and Technology, Nihon University, 1-8 Kanda-Surugadai, Chiyoda-ku,Tokyo, 101-0062, Japan}

\author{Hiroki Yamamoto}
\affiliation{Takasaki Institute for Advanced Quantum Science, National Institutes for Quantum Science and Technology (QST), 1233 Watanuki, Takasaki, Gunma, 370-1292, Japan}

\author{Keiichirou Akiba}
\affiliation{Takasaki Institute for Advanced Quantum Science, National Institutes for Quantum Science and Technology (QST), 1233 Watanuki, Takasaki, Gunma, 370-1292, Japan}

\author{Kouichi Takase}
\affiliation{College of Science and Technology, Nihon University, 1-8 Kanda-Surugadai, Chiyoda-ku, Tokyo, 101-0062, Japan}

\date{\today}

\begin{abstract}
We investigated the magnetic and electric properties of nanometer-sized vanadium dioxide (VO$_2$) 
particles. 
 VO$_2$ nanoparticles were formed by milling VO$_2$ powder. 
We measured the magnetic field dependence of the magnetization of the VO$_2$ powder and nanoparticles. 
The VO$_2$ powder did not exhibit ferromagnetism, 
whereas the VO$_2$ nanoparticles exhibited ferromagnetism.  
In addition, we fabricated samples by bridging between electrodes with the VO$_2$ nanoparticles, and
the temperature dependence of their resistance was measured.
Metal-insulator transitions (MITs) were observed, 
and the temperature range where the MIT occurred was wider than that in a typical bulk VO$_2$. 
The VO$_2$ nanoparticles exhibited these properties of ferromagnetism and MIT 
possibly because of the surface and size effects of the VO$_2$ nanoparticles. 
These results indicate the first observation of the competitive coexistence of ferromagnetism and MIT 
of VO$_2$ nanoparticles.
\end{abstract}

\maketitle

\section{Introduction}

Vanadium dioxide (VO$_2$) exhibits a metal-insulator transition (MIT) at $\sim$ 340 K
\cite{Morin,Mott,Berglund,Zylbersztejn,Pouget,Wentzcovitch,Rice}. 
The MIT is accompanied by a structural phase transition (SPT) 
from a high-temperature tetragonal metallic rutile phase to a low-temperature insulating monoclinic one, 
and the value of the electric resistivity changes by more than several orders 
of magnitude \cite{Morin,Berglund2,Kim,Yao,Qazilbash}. VO$_2$ 
is expected to be applied to novel electronic devices using MIT 
\cite{Kim2,Nakano,Liu,Ji,Jeong,Beaumont,Yajima,Shibuya,Tsuji}. 
Recently, the MIT of low-dimensional VO$_2$, such as thin films \cite{Kim3,Yang,Shiga,Barimath}, nanobeams\cite{Wu,Wei,Hong}, 
and nanoparticles \cite{Li,Beckerle}, has been studied.  
More recently, ferromagnetism has also been reported in VO$_2$ nanoparticles\cite{Fukawa}. 
Although ferromagnetism in nanoparticles may occur 
because of many defects on the surface of VO$_2$ nanoparticles\cite{Fukawa},
its detailed mechanism has not yet been completely elucidated. 
Thus, both the magnetic and electric properties of VO$_2$ nanoparticles must be analyzed. 
The magnetic properties of nanoparticles can easily be determined 
using a superconducting quantum interference device (SQUID). 
Meanwhile, one simple way to observe the MIT of VO$_2$ nanoparticles is to measure the variations of electric resistance values. 
However, measuring electric resistance by attaching electrodes to VO$_2$ nanoparticles is difficult.
Therefore, there have been no experimental results that measure ferromagnetism 
by the magnetic measurement and the MIT by the electric measurement. 

In this study, we prepared VO$_2$ nanoparticles by milling VO$_2$ powder, 
and the magnetic field dependence of the magnetization of VO$_2$ nanoparticles 
was observed using SQUID. 
Furthermore, VO$_2$ samples were fabricated by bridging between submicron-sized nanogap electrodes 
with the VO$_2$ nanoparticles. 
Then, the temperature dependence of the resistance of the VO$_2$ samples was measured. 
Finally, the relation between ferromagnetism and MIT of the VO$_2$ nanoparticles was discussed.

\section{Experimental Preparation}

\begin{figure}
\begin{center}
\includegraphics[width=1\columnwidth]{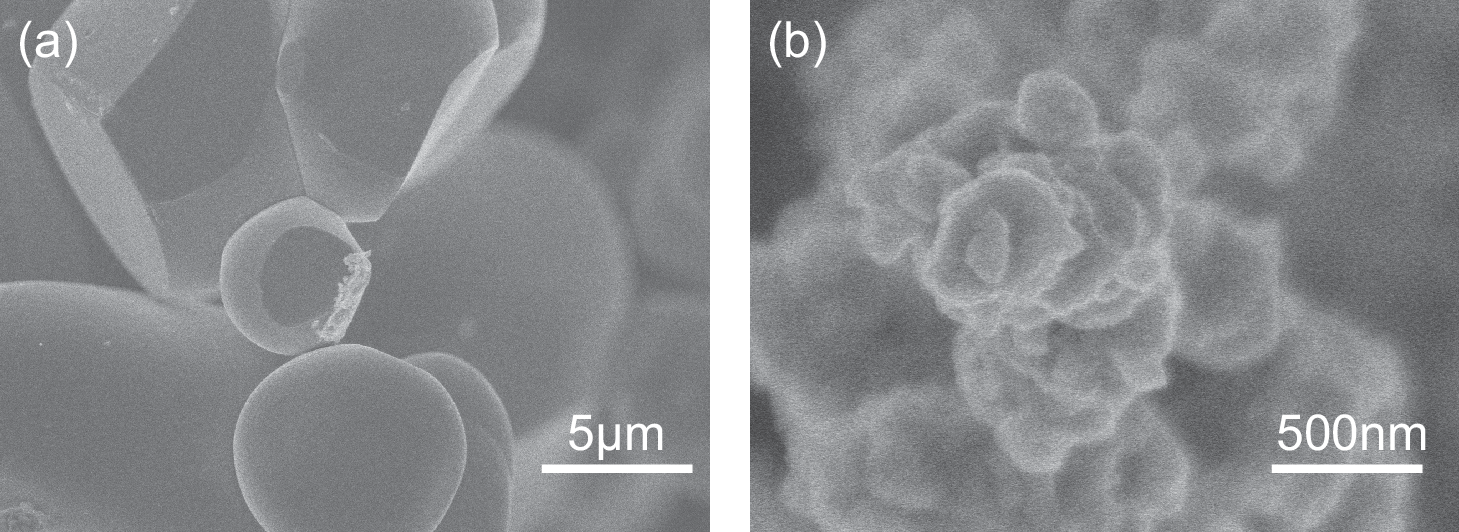}
\caption{
SEM images of (a) the VO$_2$ powder
and (b) milled VO$_2$ nanoparticles.
}
\label{psem}
\end{center}
\end{figure}

VO$_2$ nanoparticles were prepared by milling a commercial
powder product (Kojundo Chemical Lab, 99.9\%)\cite{hatano2} using ZrO$_2$ balls and a ZrO$_2$ vessel 
at  a rotation speed of 400 rpm for 1 h. 
Here, we confirmed that ZrO$_2$ shows paramagnetism and not ferromagnetism 
by measuring the magnetic field dependence 
of the magnetization of ZrO$_2$ balls\cite{Fukawa}. 
Figure \ref{psem} (a) shows a scanning electron microscope (SEM) image 
of the VO$_2$ powder. 
As shown in Fig. \ref{psem} (a), the VO$_2$ powder was composed of micrometer-sized
particles with smooth surfaces. 
Meanwhile, the SEM image of the milled VO$_2$ nanoparticles is shown in Fig. \ref{psem} (b). 
The VO$_2$ powder was crushed to nanometer-sized particles with pebble-grained surfaces.  
The average diameter of the VO$_2$ nanoparticles was estimated as 42 nm 
using the Scherrer equation\cite{Fukawa}.

\begin{figure}
\begin{center}
\includegraphics[width=1\columnwidth]{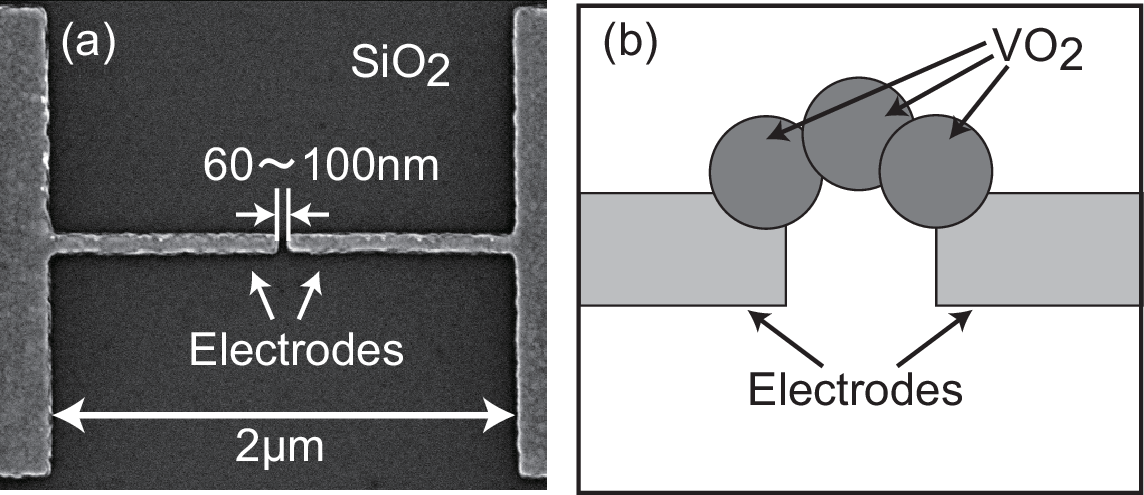}
\caption{
 (a) SEM image of nanogap electrodes. (b) Schematic of the VO$_2$ sample.
}
\label{sample}
\end{center}
\end{figure}

The VO$_2$ samples were fabricated using the milled VO$_2$ nanoparticles. 
First, a thermal oxide with a thickness of 50 nm was grown on an n-type Si substrate.  
Then, nanogap electrodes with Ti and Au were created on substrates 
using 100 kV electron beam lithography (ELIONIX, ELS-G100-SP) 
and a metal evaporator. 
Figure \ref{sample} (a) shows the SEM image of typical nanogap electrodes. 
The distances between the two electrodes were $60 - 100$ nm, 
and the thicknesses of Ti and Au were 5 and 20 nm, respectively. 
Then, the prepared VO$_2$ nanoparticles were dissolved 
in ethanol by stirring with ultrasonic waves, and a VO$_2$ solution was created. 
Next, the VO$_2$ solution was sprayed onto the SiO$_2$, 
where the nanogap electrodes were located. 
Finally, ethanol was evaporated at room temperature, and VO$_2$ samples were fabricated. 
The schematic of a VO$_2$ sample is shown in Fig. \ref{sample} (b). 
The average diameter of the VO$_2$ nanoparticles was 42 nm. 
Therefore, several nanoparticles must be located between the nanogap electrodes 
to form a current path. 
We confirmed current flows in two samples (samples A and B). 

\section{Results and Discussion}

\begin{figure}
\begin{center}
\includegraphics[width=1\columnwidth]{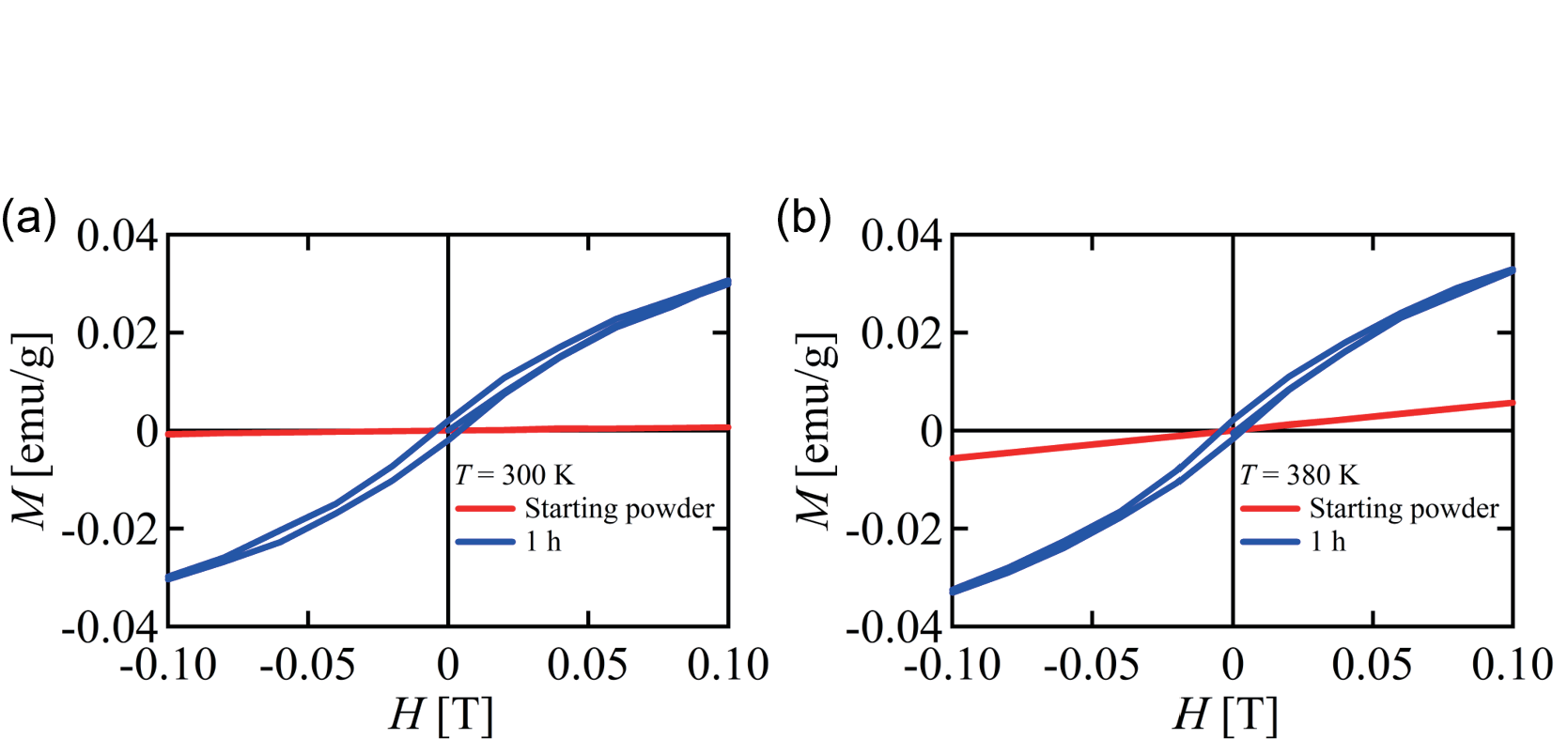}
\caption{
Magnetic field ($H$) dependence of the magnetization ($M$) of VO$_2$ powder 
and milled VO$_2$ nanoparticles at (a) $T=$ 300 K and (b) $T=$ 380 K.
}
\label{particlehm}
\end{center}
\end{figure}

We measured the magnetization ($M$) of VO$_2$ 
powder and nanoparticles as a function of the magnetic field ($H$). 
Figure \ref{particlehm} (a) shows the $H$ dependence of $M$ of the VO$_2$ powder  
and milled VO$_2$ nanoparticles at $T=$ 300 K. 
As shown in Fig. \ref{particlehm} (a), no magnetization was observed for the VO$_2$ powder.  
Meanwhile, the variation of $M$ with hysteresis was observed for the VO$_2$ nanoparticles, 
indicating ferromagnetism for the insulating phase.  
The $H$ dependence of $M$ of the VO$_2$ 
powder and nanoparticles at $T=$ 380 K is shown in Fig. \ref{particlehm} (b). 
In this figure, we observed accreted hysteresis loops for the VO$_2$ nanoparticles 
but not for the VO$_2$ powder. 
This also indicated ferromagnetism for the metallic phase in the VO$_2$ nanoparticles.  
Therefore, the VO$_2$ nanoparticles exhibited ferromagnetism independent of the crystal structure 
(i.e., insulating and metallic phases).  
This ferromagnetism may have been due to many defects on pebble-grained surface\cite{Fukawa}. 

\begin{figure}
\begin{center}
\includegraphics[width=0.5\columnwidth]{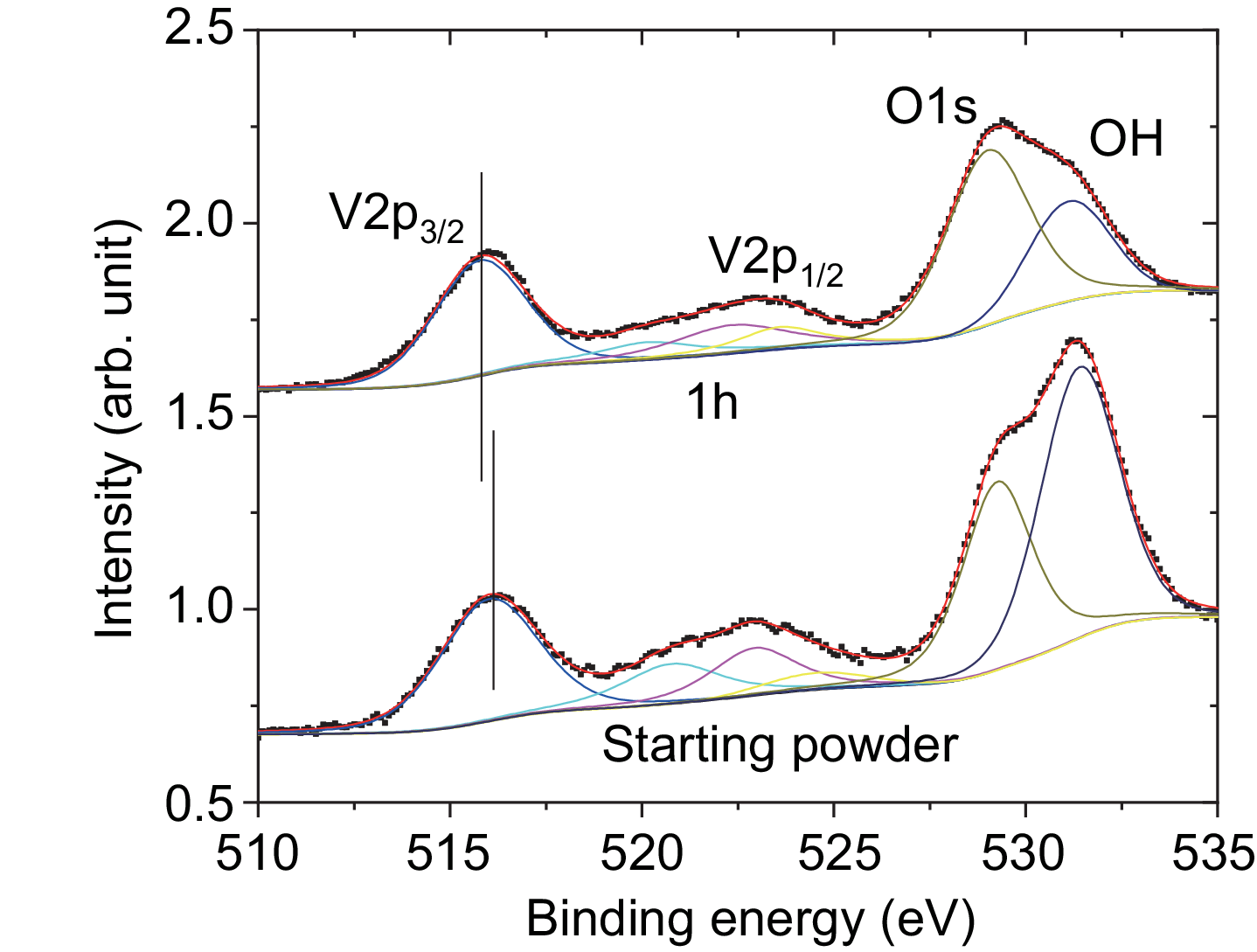}
\caption{
XPS measurement results and peak fitting of VO$_2$ starting powder and nanoparticles.
}
\label{xps}
\end{center}
\end{figure}

To investigate whether defects are formed by milling the VO$_2$ powder, 
we analyzed VO$_2$ starting powder and nanoparticles using X-ray photoelectron spectroscopy (XPS).  
Figure \ref{xps} shows the XPS spectrum and the peak fitting of the VO$_2$ starting powder 
and nanoparticles. 
The dots are the measurement results. 
The background was obtained by the Shirley method,  
and the red curves are the fits to the XPS spectrum. 
The blue, purple, green, and dark blue curves show the contributions of the V 2p$_{3/2}$, V 2p$_{1/2}$, O 1s 
and OH peaks, respectively. 
The light blue and yellow curves show the satellite peaks of V 2p$_{3/2}$ and V 2p$_{1/2}$,
which appear due to the strong hybridization between the V 2p$_{3/2}$ or  V 2p$_{1/2}$ and O 1s. 
In this figure, the V 2p$_{3/2}$ peak of VO$_2$ nanoparticles redshifts by 0.3 eV with respect to the starting powder. 
This redshift may be due to the presence of oxygen defects. 
However, the VO$_2$ starting powder is milled in air, so the VO$_2$ nanoparticles are pre-exposed to atmospheric oxygen. 
Therefore, it is difficult to unambiguously claim that oxygen defects present on the surface of VO$_2$ nanoparticles.

\begin{figure}
\begin{center}
\includegraphics[width=1\columnwidth]{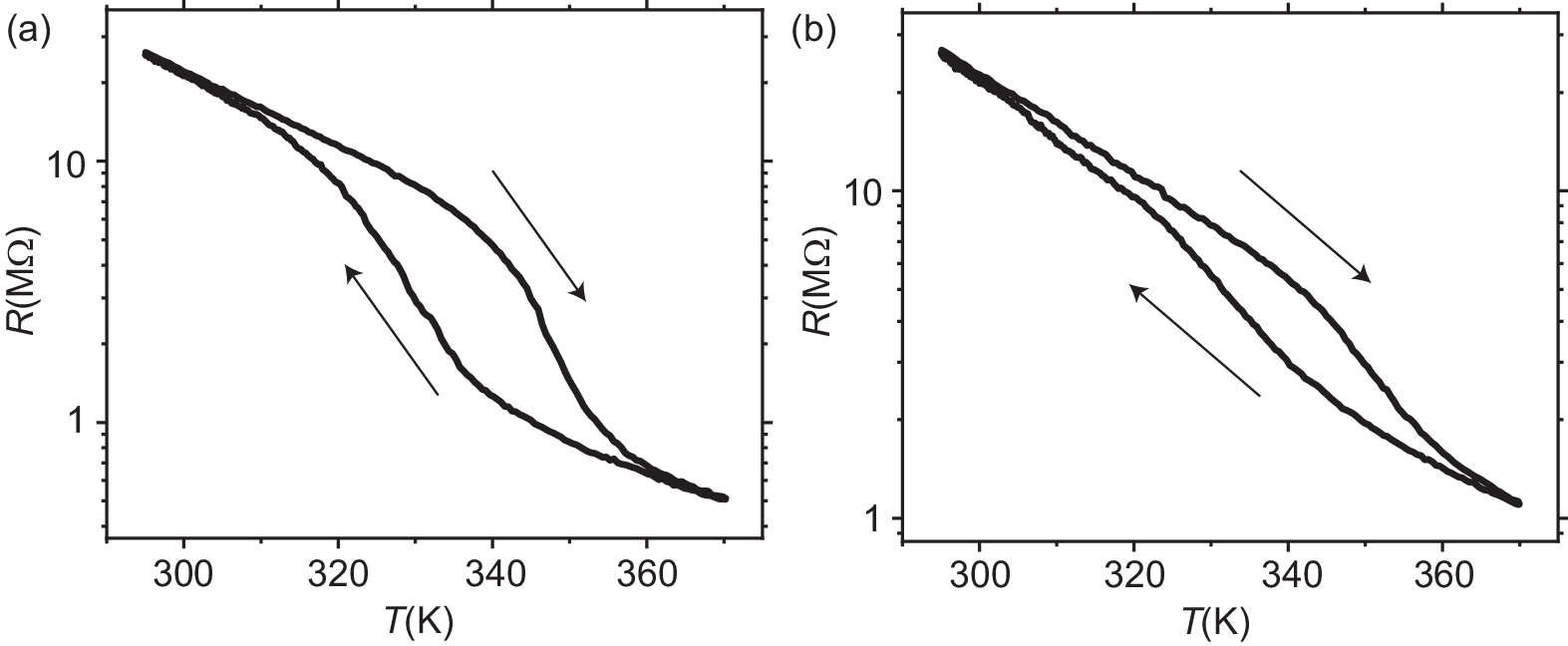}
\caption{
Temperature ($T$) dependence of resistance ($R$) in VO$_2$ samples (a) A and (b) B.
}
\label{tempdep}
\end{center}
\end{figure}

Next, we measured the resistance ($R$) of the two samples as a function of $T$. 
Here, the source-drain voltage was fixed at 100 mV.
The $T$ dependence of $R$ of sample A is shown in Fig. \ref{tempdep} (a). 
Here, the two arrows indicate the direction of the increase and decrease in $R$. 
As $T$ increased from $\sim$295 K up to $\sim$340 K, $R$ gradually decreased from $\sim$27 M$\Omega$.
Then, when $T$ exceeded $\sim$340 K, $R$ rapidly decreased.
As $T$ further increased up to $\sim$360 K, 
the value of $R$ slowly decreased again. 
Finally, the value of $R$ was $\sim$500k$\Omega$ at $T=$ 370 K, 
which was approximately 60 times lower than that at $T=$ 295 K. 
Next, when $T$ decreased from $\sim$370 K, $R$ gradually increased 
until $T$ reached approximately 340 K.
When $T$ decreased from $\sim$340 K to $\sim$320 K, $R$ rapidly increased 
and then increased slowly again until 295 K. 
The value of $R$ returned to the initial one
and the variation of $R$ showed hysteresis. 
Therefore, the MIT occurs in sample A.  
Similarly, the $T$ dependence of $R$ of sample B is shown in Fig. \ref{tempdep} (b). 
As shown in this figure, the $T$ dependence of $R$ of sample B was 
 almost the same as that of sample A. 
However, the temperature range where the MIT occurred was wider than that of sample A. 
Moreover, the minimum value of $R$ was  $\sim$1.1 M$\Omega$, 
which $\sim$30 times lower than that at $T=$ 295 K. 
For the MIT of bulk VO$_2$,  
the value of its resistance changes by more than five orders of magnitude 
in a very narrow temperature range of several K at approximately $T=$340 K 
 \cite{Morin,Berglund2,Kim,Yao,Qazilbash}.
Therefore, the $R$ of the VO$_2$ nanoparticles changed 
more slowly than that of bulk VO$_2$ as $T$ changed.

\begin{figure}
\begin{center}
\includegraphics[width=1\columnwidth]{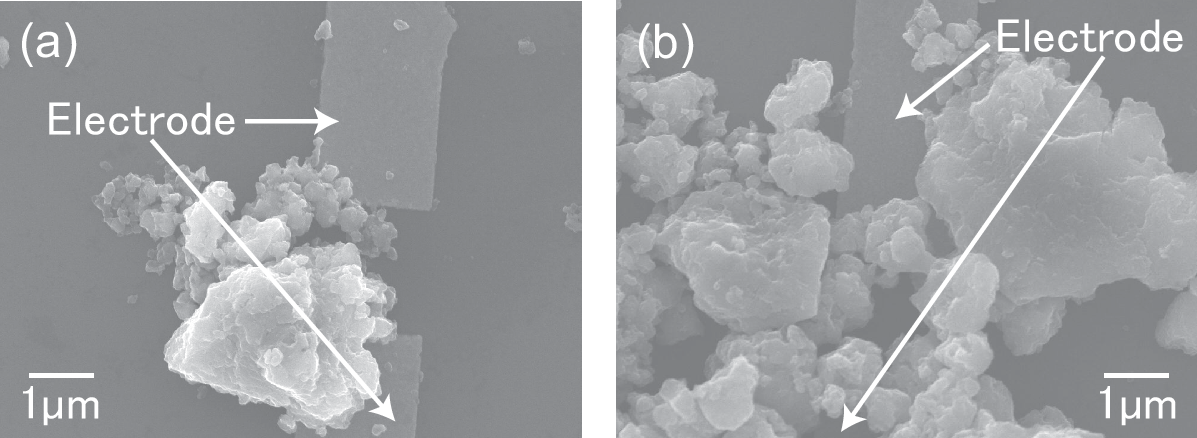}
\caption{
SEM images of VO$_2$ samples (a) A and (b) B.
}
\label{sem}
\end{center}
\end{figure}

To discuss the electric proprieties of the VO$_2$ samples in detail, 
the SEM images of samples A and B are shown in Figs. \ref{sem} (a) and (b), respectively. 
In these figures, 
some VO$_2$ nanoparticles were not located just between a spacing of $\sim$100 nm of  
two nanogap electrodes, which corresponded to the narrowest width between them. 
Instead, many VO$_2$ nanoparticles were accidentally located 
between approximately 2 ${\rm \mu}$m of the two nanogap electrodes, 
which corresponded to the second narrowest width between them as shown in Fig. \ref{sample} (a) \cite{hatano}. 
Because the inner part of the VO$_2$ nanoparticles at high temperatures is metallic, 
the values of the contact resistances between the VO$_2$ nanoparticles are higher than 
that of the inner part of the VO$_2$ nanoparticles.  
Here, the high contact resistances may be caused 
by electrons trapped in many defects on the pebble-grained surfaces. 
Therefore, the values of $R$ of the two samples were determined 
by the contact resistance between the VO$_2$ nanoparticles,  
not the resistances of the inner part of the VO$_2$ nanoparticles. 
As described above, many VO$_2$ nanoparticles were located between the two electrodes. 
Thus, the values of $R$ of the two samples were much larger than those of typical bulk VO$_2$.  
Besides, as in Fig. \ref{psem} (b), 
the diameters of most of VO$_2$ nanoparticles were lower than approximately 100 nm.  
However, some of them appeared to be aggregated to micrometer size,  
and micrometer-sized VO$_2$ particles were located between the two nanogap electrodes.  
Therefore, the two samples had different hysteresis shapes and different minimum resistance values 
maybe because the VO$_2$ particles had not only nanometer sizes 
but also dispersed sizes due to aggregation.

\section{Conclusions}

We made VO$_2$ nanoparticles by milling VO$_2$ powder 
and fabricated VO$_2$ samples using these nanoparticles. 
Ferromagnetism was observed by measuring the magnetic field dependence  
of the magnetization of VO$_2$ nanoparticles. 
Many defects on the pebble-grained surface of the VO$_2$ nanoparticles 
may have resulted in the exhibition of ferromagnetism. 
However, 
although MIT was also obtained in the temperature dependence 
of the resistance of the VO$_2$ samples, 
it was exhibited over a wider temperature range than that of bulk VO$_2$.  
This wider hysteresis of MIT may be because the VO$_2$ particles had not only nanometer sizes 
but also dispersed sizes due to aggregation.  
We must thus continue researching the reasons why ferromagnetism and 
MIT are exhibited in VO$_2$ nanoparticles.  
In conclusion, this research is the first demonstration of competitive coexistence of 
ferromagnetism and MIT in VO$_2$ nanoparticles. 
Besides, the utilization of the ferromagnetism and phase transition of VO$_2$nanoparticles, 
is expected to create novel spintronic devices, 
such as magneto-resistive sensor devices whose resistance can be controlled by temperature.

\begin{acknowledgments}


This work was supported in part by the Nihon University
Multidisciplinary Research Grant (2021), the Research Grant of
College of Engineering, Nihon University (2023), and JSPS
KAKENHI (Nos. 20H02489 and 21KK0262) from the
Ministry of Education, Culture, Sports, Science and
Technology of Japan (MEXT). Also, a part of this work was
conducted at the Institute for Molecular Science, supported by
“Advanced Research Infrastructure for Materials and
Nanotechnology in Japan (ARIM)” of the Ministry of
Education, Culture, Sports, Science and Technology (MEXT)
Proposal Number JPMXP1223OS1025.

\end{acknowledgments}


\begin{thebibliography}{99}
\bibitem{hatano2}
We confirmed the MIT of the VO$_2$ powder by measuring the temperature dependence of their magnetization.
\bibitem{hatano}
It is impossible to accurately control the condition, that  VO$_2$ nanoparticles are located, 
because the VO$_2$ nanoparticles are distributed on the nanogap electrodes using splay.
\end{thebibliography}
\end{document}